\documentclass[9pt,twocolumn,twoside]{osajnl}

\journal{ol} 

\setboolean{shortarticle}{true} 

\ifthenelse{\boolean{shortarticle}}{\colorlet{color2}{color2b}}{\colorlet{color2}{color2}} 

\title{Effect of dark counts on single photon heralding with quasi-number-resolving detection schemes}

\author{L. G. Helt}
\author{M. J. Steel}

\affil{Centre for Ultrahigh bandwidth Devices for Optical Systems (CUDOS), MQ Photonics Research Centre, QSciTech Research Centre, Department of Physics and Astronomy, Macquarie University, NSW 2109, Australia}

\dates{Compiled \today}

\ociscodes{(270.5290) Photon statistics; (270.5570) Quantum detectors; (040.1435) Avalanche photodiodes (APDs).}

\doi{\url{http://dx.doi.org/10.1364/OL.XX.XXXXXX}}

\begin{abstract}
We consider photon heralding with quasi-number-resolving detection schemes and account for detection efficiencies and dark count probabilities. With a straightforward formalism, we develop closed-form expressions for the heralding probability, photon number distribution of the resulting heralded state, and fidelity of this heralded state to a single photon state. We calculate that, on the basis of optimizing this fidelity, due to the presence of dark counts there is a maximum number of detection modes worth multiplexing for each combination of efficiency and dark count probability.
\end{abstract}

\setboolean{displaycopyright}{true}

\begin{document}

\maketitle
\thispagestyle{fancy}

\ifthenelse{\boolean{shortarticle}}{\ifthenelse{\boolean{singlecolumn}}{\abscontentformatted}{\abscontent}}{}

The ability to generate, interfere, and detect individual photons is at the heart of optical implementations of quantum information processing applications. These applications include communications, metrology, and computing that promise quantum enhancements compared to their classical counterparts~\cite{Lee:2005}. While the field continues to progress experimentally~{\cite{Sibson:2017, Whittaker:2017, Wang:2016}, it is hampered by the current difficulty in obtaining large numbers of identical single photons.

To date, the most common method of producing single photons has been via probabilistic or ``heralded'' sources, in which a pair of photons is stochastically generated and one member of the pair detected to indicate the presence of the other~\cite{Clark:2015}. These sources make use of the nonlinear optical processes of either spontaneous parametric downconversion, or spontaneous four-wave mixing, to probabilistically convert bright classical pump pulse photons into photon pairs. Taking advantage of waveguide engineering, multiple heralded sources can be made to produce near-identical single photons~\cite{Spring:2017}. Unfortunately, the probabilistic nature of these sources means that the greater the number of identical single photons simultaneously desired, the lower the rate at which they can be generated.

For a given pump pulse incident on a heralded single photon source, there is an associated photon-pair number distribution that describes the state of generated photons before detection.  In typical sources, incident pump pulse powers are kept low enough to ensure that this distribution is dominated by the zero-pair and single-pair terms.  This, in turn, allows common on-off detectors (e.g. silicon avalanche photodiodes), only able to distinguish between zero and nonzero incident photons, to herald single photons with reasonable veracity.  In this regime, the single-pair probability is typically less than 1\%. However, at higher pump powers the single-pair probability can be made as high as 25\%~\cite{Christ:2012}, and thus single photons could be heralded with much higher probabilities provided the detection scheme employed were capable of distinguishing between zero, one, or more than one incident photon.

Number-resolving detectors, such as transition-edge sensors~\cite{Lita:2008}, provide one option, but are complex, expensive, and require significant overhead to operate. Constrained to work with on-off detectors only, an alternate option is known as a quasi-number-resolving (QNR) detection scheme. These schemes involve splitting the desired detection mode across several spatial~\cite{Paul:1996, Sperling:2012, Heilmann:2016, Matthews:2016, Spring:2017} or temporal~\cite{Rehacek:2003, Fitch:2003, Achilles:2003, Achilles:2004, Kruse:2017} modes before detection via an on-off detector. Split enough times, each individual detection need only distinguish between zero or nonzero photons for accurate number resolution. However, previous theoretical works on QNR detection schemes appear to have largely ignored trade-offs that come with the application to single photon heralding as well as, with the exception of~\cite{Sperling:2012}, the impact of detector dark counts.

In this Letter we consider a heralded single photon source employing a QNR detection scheme in the heralding mode. We develop simple, closed-form expressions for the heralding mode detection probability, the photon number distribution of the heralded state, and the fidelity of this state to a single photon state, accounting for detector efficiencies and dark counts. We show that, in contrast to the intuitive viewpoint that adding more detectors is always beneficial~\cite{Heilmann:2016}, the presence of dark counts means that that there is always a point beyond which increasing the number of detectors is detrimental. 

While, in general, the state of generated photons produced by an ultrafast pump pulse incident on a modern (waveguide) source can contain frequency correlations between photons in heralding and heralded modes, we note that such correlations are known to limit the fidelities of heralded states and as such considerable work has been done on the engineering of these devices to remove such correlations~\cite{Grice:2001, Konig:2004, Branczyk:2010, Branczyk:2011, Zhang:2012, Dixon:2013, Harder:2013, Dosseva:2016, Tambasco:2016, Spring:2017}. Consequently, we focus our attention here on sources that are free of frequency correlations between heralding and heralded modes, reserving the inclusion of multi-spectral-mode effects for future work.

We begin our calculation with the pre-heralding two-mode state produced per pump pulse by a heralded source
\begin{equation}
\label{eq:sqvac}
\left\vert\psi\right\rangle=\sum_n c_n \left\vert n\right\rangle_A \left\vert n\right\rangle_B,
\end{equation}
characterized by the thermal distribution
\begin{equation}
\label{eq:thermal}
\left\vert c_{n}\right\vert^{2}=\frac{\mu^{n}}{\left(1+\mu\right)^{1+n}},
\end{equation}
where $\mu$ is the average number of pairs generated per pump pulse and tensor products are left implicit. Operating as a heralded source, detection in mode $A$ will herald the state in mode $B$. As we ultimately wish to consider the splitting of mode $A$ equally across many modes, we note that split across the two modes $A_{1}$ and $A_{2}$ with an appropriate spatial or temporal beam splitter, we can write this state as
\begin{equation}
\left|\psi\right\rangle \rightarrow\left|\psi_{2}\right\rangle=\sum_{n}c_{n}\sqrt{\frac{n!}{2^{n}}}\sum_{k_{i}\vert k_{1}+k_{2}=n}\frac{1}{\sqrt{k_{1}!k_{2}!}}\left|k_{1}\right\rangle _{A_{1}}\left|k_{2}\right\rangle _{A_{2}}\left|n\right\rangle _{B},
\end{equation}
where the notation $k_{i}\vert k_{1}+k_{2}=n$ means summing over all possible $k_{1}$ and $k_{2}$ satisfying $k_{1}+k_{2}=n$. Here the $2^{n}$ comes from equal mode splitting, and the factorials from bosonic operators and binomial coefficients. This allows easy generalization to the equal splitting across $M$ modes,
\begin{align}
\label{eq:sqvacM}
\left|\psi_{M}\right\rangle=&\sum_{n}c_{n}\sqrt{\frac{n!}{M^{n}}} \nonumber \\
&\times\sum_{k_{i}\vert k_{1}+\cdots+k_{M}=n}\frac{1}{\sqrt{k_{1}!\cdots k_{M}!}}\left|k_{1}\right\rangle _{A_{1}}\cdots\left|k_{M}\right\rangle _{A_{M}}\left|n\right\rangle _{B},
\end{align}
accomplished, for example, via a binary tree of 50/50 splitters (see Fig.~\ref{fig:splitting}), in which case $M=2^{L}$, where $L$ is the number of layers of the tree, or via an appropriate linear optical network~\cite{Reck:1994}.
\begin{figure}[htbp]
\centering
\includegraphics[width=0.9\linewidth]{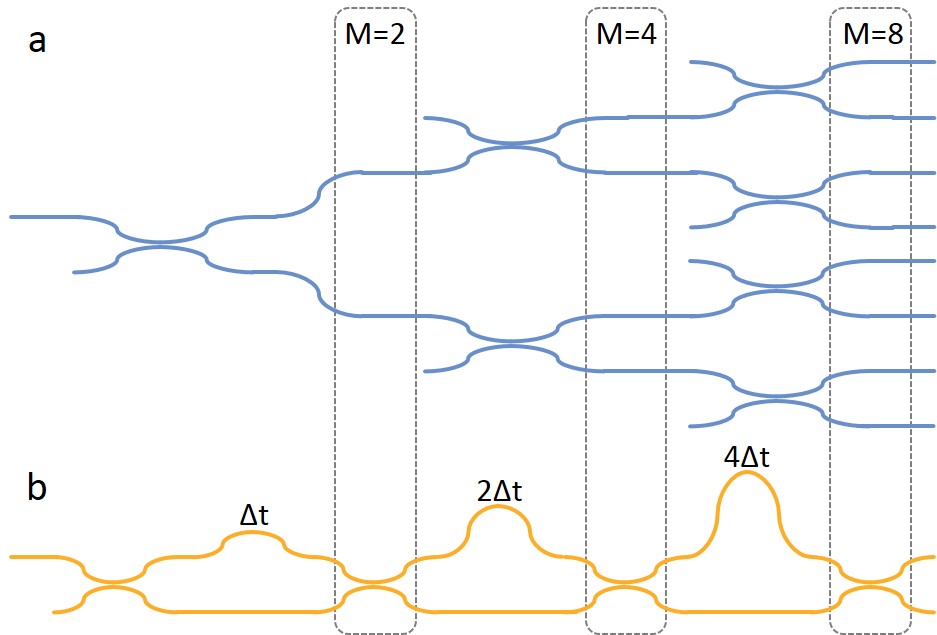}
\caption{Sketch of a scheme to split photons equally across several a. spatial or b. temporal modes.}
\label{fig:splitting}
\end{figure}

We represent on-off detection in mode $X$, with the POVM elements
\begin{equation}
\hat{\pi}_{0;X}=\left(1-\delta_{X}\right)\sum_{n}\left(1-\eta_{X}\right)^{n}\left\vert n\right\rangle_{X}\left\langle n\right\vert_{X},
\end{equation}
and
\begin{equation}
\hat{\pi}_{\text{click};X}=\hat{\mathbb{I}}_{X}-\hat{\pi}_{0;X}=\hat{\mathbb{I}}_{X}-\left(1-\delta_{X}\right)\sum_{n}\left(1-\eta_{X}\right)^{n}\left\vert n\right\rangle_{X}\left\langle n\right\vert_{X}.
\end{equation}
Here $\eta_{X}$ is the detection efficiency parameter and $\delta_{X}$ the probability per detection window (synchronized to pump pulses) of a dark count, assuming $\delta_{X}\ll 1$.  One way of interpreting these equations is to recognize $\left(1-\eta_{X}\right)^{n}\left|n\right\rangle _{X}\left\langle n\right|_{X}$ as representing the detector registering ``off'' for $n$ incident photons, the sum over $n$ as accounting for all such terms for a state represented in the number basis, and $\left(1-\delta_{X}\right)$ as reducing this value slightly due to the presence of dark counts. Propagation and coupling losses relevant to a particular experimental realization can be incorporated in $\eta_{X}$, as this model approximates an inefficient detector as a perfect detector with a lossy element placed before it. For the detector(s) to register $m$ ``on'' values, or ``clicks'' across $M$ modes, we therefore define the operator
\begin{equation}
\label{eq:QNR}
\hat{\pi}_{m;X}^{M}=\binom{N}{m}\prod_{i=1}^{m}\hat{\pi}_{\text{click};X_{i}}\prod_{j=m+1}^{M}\hat{\pi}_{0;X_{j}},
\end{equation}
where the binomial coefficient appears because it does not matter which set of $m$ modes the clicks are registered in.

Armed with these expressions, we may now calculate the probability of registering $m$ clicks associated with photons originally in heralding mode $A$, and described by the state in~\eqref{eq:sqvac}, with the QNR detection scheme of Eqs.~(\ref{eq:sqvacM}) and~(\ref{eq:QNR})
\begin{equation}
P_{A}^{M}\left(m\right)=\text{Tr}\left(\hat{\pi}_{m;A}^{M}\left\vert\psi_{M}\right\rangle\left\langle\psi_{M}\right\vert\right).
\end{equation}
We can also calculate the photon number distribution of the associated heralded state
\begin{equation}
\rho_{m;B}^{M}=\frac{\text{Tr}_{A}\left(\hat{\pi}_{m;A}^{M}\left\vert\psi_{M}\right\rangle\left\langle\psi_{M}\right\vert\right)}{P_{A}^{M}\left(m\right)},
\end{equation}
and the fidelity of the heralded state to an $\ell$-photon Fock state
\begin{equation}
\label{eq:F}
F^{M}\left(\ell\vert m\right)=\left\langle\ell\right\vert_{B}\rho_{m;B}^{M}\left\vert\ell\right\rangle_{B}.
\end{equation}
In particular, setting $\delta_{A_{i}}=\delta$ and $\eta_{A_{i}}=\eta$ for all $i$, we find
\begin{equation}
P_{A}^{M}\left(m\right)=\sum_{n}\left|c_{n}\right|^{2}\binom{M}{m}\sum_{j=0}^{m}\left(-1\right)^{m-j}\binom{m}{j}\left(1-\delta\right)^{M-j}\left[\left(1-\eta\right)+\frac{j\eta}{M}\right]^{n}.
\end{equation}
We note that the zero dark count probability limit of this expression was previously derived by different methods in~\cite{Fitch:2003}, and the zero dark count as well as perfect detection efficiency limit
\begin{align}
P_{A}^{M}\left(m\right)_{\delta=0,\eta=1}=&\sum_{n=m}^{\infty}\left|c_{n}\right|^{2}\frac{M!M^{-n}}{m!\left(M-m\right)!}\left(\sum_{j=0}^{m}\frac{m!\left(-1\right)^{j}\left(m-j\right)^{n}}{j!\left(m-j\right)!}\right)\nonumber\\
=&\sum_{n=m}^{\infty}\left|c_{n}\right|^{2}\frac{M!S\left(n,m\right)}{\left(M-m\right)!M^{n}},
\end{align}
 has been previously derived by different methods in each of~\cite{Paul:1996, Sperling:2012, Matthews:2016}. Here
\begin{equation}
S\left(n,m\right)=\frac{1}{m!}\sum_{j=0}^{m}\left(-1\right)^{m-j}\binom{m}{j}j^{n},
\end{equation}
is known as the Stirling number of the second kind, and counts the number of ways to partition a set of $n$ objects, in our case, photons, into $m$ non-empty subsets, in our case, detection modes. Making use of~\eqref{eq:thermal} and turning our attention to heralding a single photon, we arrive at the closed-form expressions
\begin{equation}
\label{eq:P}
P_{A}^{M}\left(1\right)=\xi\left[\frac{1}{1+\eta\mu\left(1-\frac{1}{M}\right)}-\frac{1-\delta}{1+\eta\mu}\right],
\end{equation}
\begin{align}
\rho_{1;B}^{M}=&\frac{\sum_{n}\left|c_{n}\right|^{2}\xi\left\{ \left[\left(1-\eta\right)+\frac{\eta}{M}\right]^{n}-\left(1-\delta\right)\left(1-\eta\right)^{n}\right\}\left|n\right\rangle _{B}\left\langle n\right|_{B}}{P_{A}^{M}\left(1\right)}\nonumber\\
=&\frac{\sum_{n}\frac{\mu^{n}}{\left(1+\mu\right)^{1+n}}\left\{ \left[\left(1-\eta\right)+\frac{\eta}{M}\right]^{n}-\left(1-\delta\right)\left(1-\eta\right)^{n}\right\}\left|n\right\rangle _{B}\left\langle n\right|_{B}}{\frac{1}{1+\eta\mu\left(1-\frac{1}{M}\right)}-\frac{1-\delta}{1+\eta\mu}},
\end{align}
and
\begin{equation}
\label{eq:FM}
F^{M}\left(1\vert 1\right)=\frac{\left|c_{1}\right|^{2}\xi\left[\frac{\eta}{M}+\delta\left(1-\eta\right)\right]}{P_{A}^{M}\left(1\right)}=\frac{\frac{\mu}{\left(1+\mu\right)^{2}}\left[\frac{\eta}{M}+\delta\left(1-\eta\right)\right]}{\frac{1}{1+\eta\mu\left(1-\frac{1}{M}\right)}-\frac{1-\delta}{1+\eta\mu}},
\end{equation}
where
\begin{equation}
\label{eq:xi}
\xi=M\left(1-\delta\right)^{M-1},
\end{equation}
is a term that drops out of both $\rho_{1;B}^{M}$ and $F^{M}\left(1\vert 1\right)$.
While it is true that, for negligible dark count probabilities, these expressions approach those associated with true number-resolving detection schemes as $M\rightarrow\infty$, the inclusion of dark counts breaks this intuitive connection, as we explore below in detail.

The potential benefits of single photon heralding with a QNR scheme compared to a single on-off detector can be seen by comparing the photon number distributions associated with photonic states both before and after heralding. Before heralding, the single-pair component of each generated state $\left\vert c_{1}\right\vert^{2}$ is largest for $\mu=1$ [recall~\eqref{eq:thermal}], and therefore here and throughout the rest of this work we consider a heralded source operating at this point. The photon number distribution of interest prior to heralding is simply $\left\vert c_{n}\right\vert^{2}$. After heralding, it is given by the overlap of the heralded state with an $n$-photon state, or $F^{M}\left(n\vert 1\right)$ as defined in~\eqref{eq:F}. Taking $\eta=0.8$ and $\delta=0.0005$ as representative detector parameters, we plot $\left\vert c_{n}\right\vert^{2}$ as well as $F^{1}\left(n\vert 1\right)$, $F^{4}\left(n\vert 1\right)$, and $F^{8}\left(n\vert 1\right)$ in Fig.~\ref{fig:distributions}. Note that $\left\vert c_{1}\right\vert^{2}=0.25$ for $\mu=1$, and, for our representative detector parameters, the single-photon component of the heralded state number distribution rises to $F^{1}\left(1\vert 1\right)\approx 0.45$, $F^{4}\left(1\vert 1\right)\approx 0.72$, or $F^{8}\left(1\vert 1\right)\approx 0.76$. While a single detector can discriminate between zero and nonzero incident photons, a QNR detection scheme can begin to discriminate higher numbers of incident photons. The slight increase in the zero-photon term as the number of modes increases in Fig.~\ref{fig:distributions} is due entirely to the presence of dark counts.
\begin{figure}[htbp]
\centering
\includegraphics[width=0.9\linewidth]{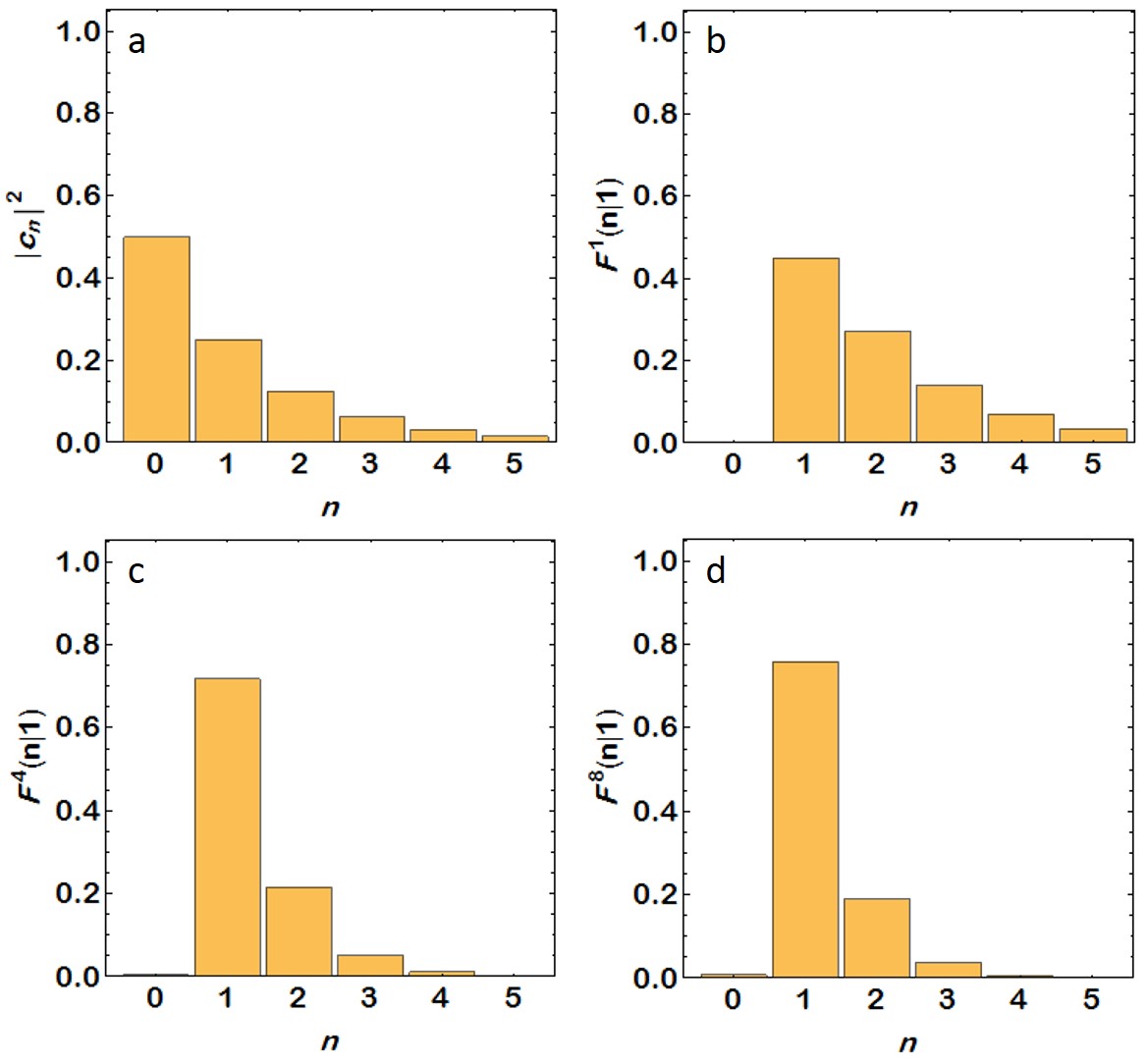}
\caption{Photon number distribution in the heralded arm of a heralded photon source a. before heralding, b. after heralding with a single on-off detector, c. after heralding with a four-mode QNR detection scheme, and d. after heralding with an 8-mode QNR detection scheme. All detectors are taken to be characterized by the representative parameters $\eta=0.8$ and $\delta=0.0005$.}
\label{fig:distributions}
\end{figure}
The probability of registering a single click [recall~\eqref{eq:P}] to herald these distributions decreases over the same range of $M$, apparently heading towards the value associated with a perfect number-resolving detector of 0.25. In particular, $P_{A}^{1}\left(1\right)\approx 0.44$, $P_{A}^{4}\left(1\right)\approx 0.29$, and $P_{A}^{8}\left(1\right)\approx 0.26$. Again, this would be expected based on the intuitive picture that adding more modes in a QND detection scheme leads to increased photon number resolving power. 

As the number of modes $M$ continues to grow, though, the behavior of both the probability of registering a single click $P_{A}^{M}\left(1\right)$, as well as the fidelity of the heralded state to a single-photon state $F^{M}\left(1|1\right)$, become less obvious. For $\mu=1$, $\eta=0.8$ and $\delta=0.0005$, after initially falling to near 0.25, $P_{A}^{M}\left(1\right)$ rises to a local maximum before falling to zero (see Fig.~\ref{fig:bigM}). The large $M$ behavior is understood simply from~\eqref{eq:xi}. As $M\rightarrow\infty$ it becomes increasingly unlikely that only a single click will be registered. Assuming that the local maximum occurs when half the single click probability is due to dark counts, taking them to be given by $1-\left(1-\delta\right)^{M}\approx M\delta$ in the limit of small $\delta$ suggests that the local maximum occurs near $M\approx 1/\left(2\delta\right)$. To obtain a better approximation, we treat $M$ as continuous in~\eqref{eq:P} and Taylor expand about $M=1/\left(2\delta\right)$ before subsequently expanding about $\delta=0$ and differentiating, to find that the local maximum occurs for
\begin{equation}
\label{eq:MP}
M^{P}_{\text{local max}}\approx\frac{5-\eta\mu}{2\delta\left(3+\eta\mu\right)},
\end{equation}
or approximately 1102 modes for our chosen parameters.  
\begin{figure}[htbp]
\centering
\includegraphics[width=0.9\linewidth]{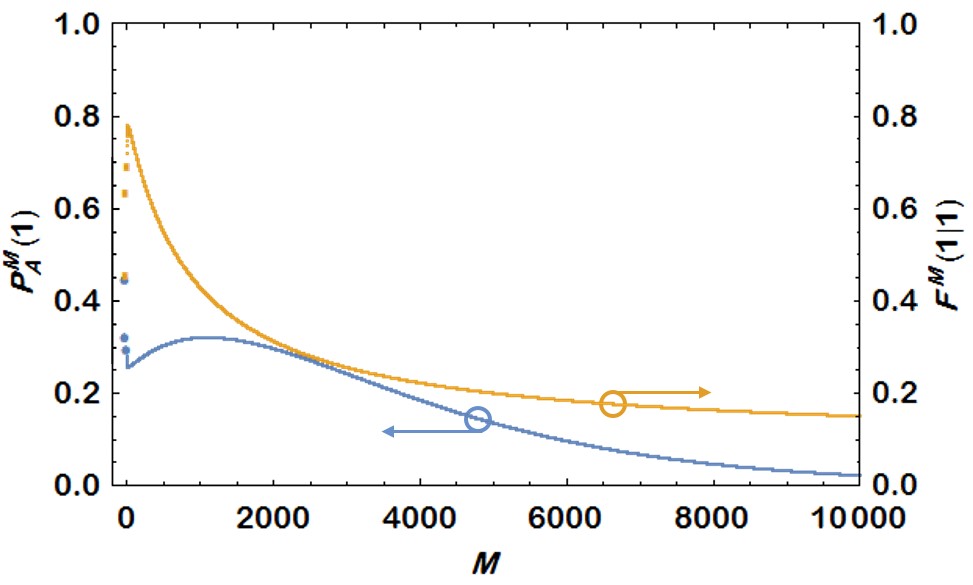}
\caption{Heralding probability and associated fidelity of the resulting heralded state to a single photon state as functions of the number of QNR detection modes $M$ for the representative values of $\eta=0.8$ and $\delta=0.0005$. The first three points of each have been made larger for greater visibility.}
\label{fig:bigM}
\end{figure}
Similarly, for any $\mu$, $\eta$, and $\delta$ there will always be an $M$ for which the fidelity $F^{M}\left(1|1\right)$ is maximal before it falls to the constant value of  $\mu\left(1+\eta\mu\right)\left(1-\eta\right)/\left(1+\mu\right)^{2}$ as $M\rightarrow\infty$ (see Fig.~\ref{fig:bigM}). This limit is easily seen from~\eqref{eq:FM} as $M\rightarrow\infty$ and the likelihood that the single registered click is due to a dark count dominates the likelihood that it is due to a photon. While curious, we note that this value is of little practical interest, as $P_{A}^{M}\left(1\right)$ approaches $0$ in the same limit, and thus the state corresponding to such a single photon fidelity would essentially never be heralded. To approximate where the maximum fidelity occurs, on the other hand, we do not need to make any Taylor expansions but simply take $M$ to be continuous and differentiate~\eqref{eq:FM} with respect to $M$ to find that
\begin{equation}
\label{eq:MF}
M^{F}_{\text{max}}\approx \frac{\eta\mu}{1+\left(2\eta-1\right)\mu}\left(1+\sqrt{\frac{1+\left[2\eta-1-\left(\eta-1\right)\delta\right]\mu}{\left(1+\eta\mu\right)\delta}}\right),
\end{equation}
provided $\delta>0$. For our parameters, the fidelity reaches a maximum with $M=22$ modes of $F^{22}\left(1|1\right)\approx 0.78$. Note that the fact that this is far less than 1105 can easily be understood by comparing the $1/\delta$ behavior of~\eqref{eq:MP} with the $1/\sqrt{\delta}$ behavior of~\eqref{eq:MF} in the $\delta\ll 1$ limit. This suggests that $F^{M}\left(1|1\right)$ will always be falling at values of $M$ for which $P^{M}_{A}\left(1\right)$ is rising, and thus the fidelity is what will most often limit how many modes are worthwhile adding. In Fig.~\ref{fig:contours}, we plot $M^{F}_{\text{max}}$ for $\mu=1$ as a series of contours, suggesting that, at least from the point of view of fidelity of the heralded state to a single photon state, there is a point beyond which continuing to increase the number of QNR detection modes $M$ has a detrimental effect for each combination of $\eta$ and $\delta$.
\begin{figure}[htbp]
\centering
\includegraphics[width=0.8\linewidth]{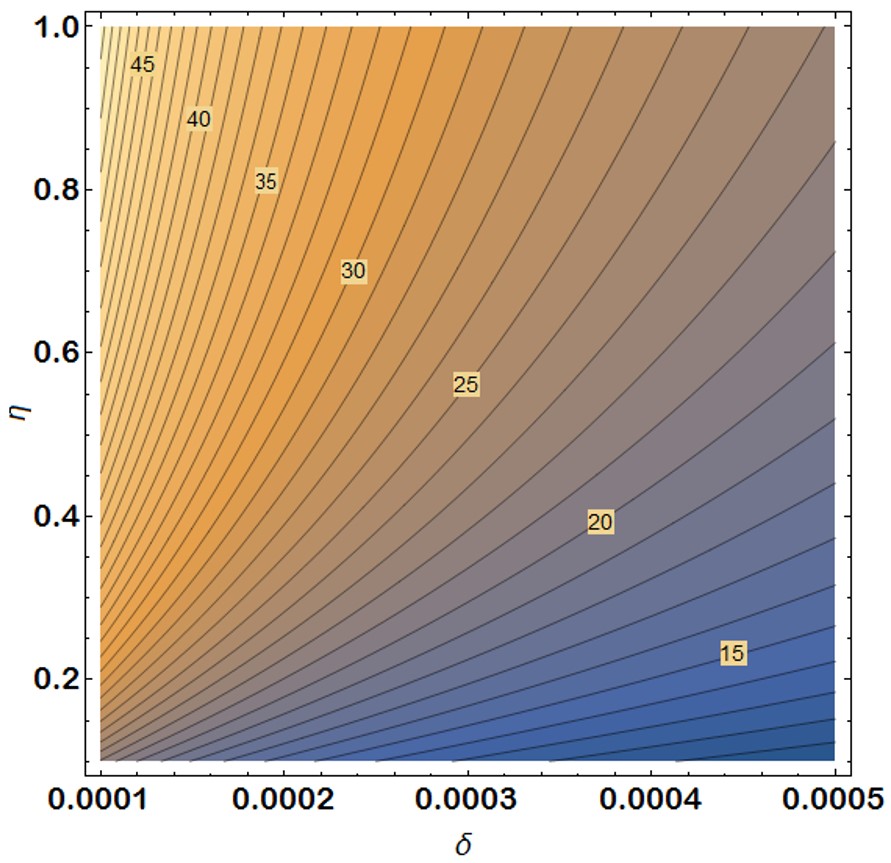}
\caption{Contours representing the approximate maximum number of QNR detection modes $M$, in the heralding arm of a heralded photon source with $\mu=1$, beyond which the fidelity of the heralded state to a single photon state only decreases.}
\label{fig:contours}
\end{figure}

In conclusion, we have given an overview of the power of QNR detection schemes for heralded single photon sources. Our simple and intuitive derivations led to closed form expressions for the  heralding mode single click probability, the photon number distribution of the heralded state, and the fidelity of this state to a single photon state. The impact of detector dark counts was taken into account and shown to limit the number of modes that are worth using for QNR heralding.  We expect these results to be useful to anyone designing a QNR detection scheme, wishing to weigh the pros and cons of adding an additional mode, and to help spur further experimental progress on photon heralding.

This research was supported by the ARC Centre for Ultrahigh bandwidth Devices for Optical Systems (CUDOS) (project number CE110001018).

\bibliography{QNR}

\ifthenelse{\equal{\journalref}{ol}}{%
\clearpage
\bibliographyfullrefs{QNR}
}{}
 
\end{document}